\documentclass[aps,prd,preprint,showpacs,superscriptaddress,nofootinbib]{revtex4}
\usepackage{slashed}
\usepackage{subfigure}
\usepackage{amsfonts,graphicx,bm}
\begin{document}

\title{Testing the Realistic Seesaw Model with Two Heavy Majorana Neutrinos
at the CERN Large Hadron Collider}

\author{Wei Chao}  \email{chaowei@ihep.ac.cn}
\affiliation{Institute of High Energy Physics, Chinese Academy of
Sciences, Beijing 100049, China}

\author{Zong-guo Si}  \email{zgsi@sdu.edu.cn}
\affiliation{Department of Physics, Shandong University, Jinan,
Shandong 250100, China}

\author{Ya-juan Zheng}  \email{yjzheng@mail.sdu.edu.cn}
\affiliation{Department of Physics, Shandong University, Jinan,
Shandong 250100, China}

\author{Shun Zhou}  \email{zhoush@ihep.ac.cn}
\affiliation{Institute of High Energy Physics, Chinese Academy of
Sciences, Beijing 100049, China}

\date{\today}

\begin{abstract}
In the conventional type-(I+II) seesaw model, the effective mass
matrix of three known light neutrinos is given by $M^{}_\nu =
M^{}_{\rm L} - M^{}_{\rm D} M^{-1}_{\rm R} M^T_{\rm D}$ in the
leading-order approximation. We propose an intriguing scenario, in
which the structural cancellation condition $M^{}_{\rm D}
M^{-1}_{\rm R} M^T_{\rm D} = {\bf 0}$ is guaranteed by the ${\rm
A^{}_4 \times Z^{}_2}$ flavor symmetry. As a consequence, neutrino
masses are mainly generated by the Higgs triplet $M^{}_\nu =
M^{}_{\rm L}$, while the neutrino mixing matrix is non-unitary and
takes on the nearly tri-bimaximal pattern. A discriminating feature
of this scenario from the pure type-II seesaw model is that the
lepton-number-violating signatures induced by the heavy Majorana
neutrinos can be discovered at the CERN Large Hadron Collider. We
calculate the total cross section of the same-sign dilepton events
$pp \to l^\pm_\alpha N^{}_i \to l^\pm_\alpha l^\pm_\beta jj$ (for
$i=1, 2$ and $\alpha, \beta = e, \mu, \tau$), and emphasize the
significant interference of the contributions from two different
heavy Majorana neutrinos. The background from the standard model and
the kinematic cuts used to reduce it have been considered. The
possible way to distinguish between the signals from heavy Majorana
neutrinos and those from doubly-charged Higgs bosons is briefly
discussed.
\end{abstract}

\pacs{14.60.St, 14.60.Pq, 14.80.Cp, 13.85.Qk}

\maketitle

\section{Introduction}

Recent neutrino oscillation experiments have provided us with very
convincing evidence that neutrinos are indeed massive and lepton
flavors do mix \cite{Review}. This great discovery implies that the
Standard Model (SM) of elementary particle physics is actually
incomplete. In order to accommodate tiny neutrino masses, one can
naturally extend the SM by introducing three right-handed neutrinos,
which are singlets under the ${\rm SU(2)^{}_{\rm L} \times
U(1)^{}_{\rm Y}}$ gauge group. In this case, the gauge invariance
allows right-handed singlet neutrinos to have a Majorana mass
$M^{}_{\rm R}$, whose scale is not subject to the gauge symmetry
breaking and then can be much larger than the electroweak scale
$\Lambda^{}_{\rm EW} \sim 10^2~{\rm GeV}$, e.g. ${\cal O}(M^{}_{\rm
R}) \sim 10^{14}~{\rm GeV} \gg \Lambda^{}_{\rm EW}$. Therefore, the
effective mass matrix for three known light neutrinos is given by
$M^{}_\nu = - M^{}_{\rm D} M^{-1}_{\rm R} M^T_{\rm D}$ in the
leading-order approximation, with $M^{}_{\rm D}$ being the Dirac
neutrino mass matrix and ${\cal O}(M^{}_{\rm D}) \sim
\Lambda^{}_{\rm EW}$. The smallness of neutrino masses can then be
ascribed to the largeness of heavy Majorana neutrino masses. This is
the so-called canonical seesaw mechanism \cite{SS1}, which offers an
elegant way to explain the tiny neutrino masses.

However, the canonical seesaw model faces two serious difficulties.
First, heavy Majorana neutrinos are too heavy and their interactions
are too weak for them to be generated in collider experiments,
especially in the forthcoming CERN Large Hadron Collider (LHC). Thus
the canonical seesaw model may lose the experimental testability.
Second, the ultrahigh energy scale characterized by the masses of
right-handed neutrinos will cause the seesaw hierarchy problem
unless ${\cal O}(M^{}_{\rm R}) \lesssim 10^7~{\rm GeV}$
\cite{SShierarchy}. One way out of these tight corners is just to
lower the seesaw scale down to TeV, but make the charged-current
interactions of heavy Majorana neutrinos sizable. This possibility
can be realized if and only if the structural cancellation condition
$M^{}_{\rm D} M^{-1}_{\rm R} M^T_{\rm D} = {\bf 0}$ is fulfilled
\cite{Pilaftsis,Smirnov}. The clear signatures of heavy Majorana
neutrinos at the LHC are then the same-sign diplepton events $pp \to
l^\pm_\alpha l^\pm_\beta jj$ (for $\alpha, \beta = e, \mu, \tau$)
\cite{Senjanovic}. Apart from heavy Majorana neutrinos, the tiny
neutrino masses can be attributed to the Higgs triplet $\Delta$,
which couples to two lepton doublets and acquires a small vacuum
expectation value $v^{}_\Delta = \langle \Delta \rangle$ \cite{SS2}.
In this case, the mass matrix of light neutrinos is given by
$M^{}_\nu = M^{}_{\rm L} = Y^{}_\Delta v^{}_\Delta$, where
$Y^{}_\Delta$ is the triplet Yukawa coupling matrix. The more
general case is the type-(I+II) seesaw model, in which both heavy
Majorana neutrinos and the Higgs triplet are present and equally
contribute to light neutrino masses. Consequently, we obtain
$M^{}_\nu = M^{}_{\rm L} - M^{}_{\rm D} M^{-1}_{\rm R} M^T_{\rm D}$.
The interplay between the terms $M^{}_{\rm L}$ and $M^{}_{\rm D}
M^{-1}_{\rm R} M^T_{\rm D}$ in the conventional type-(I+II) seesaw
model has been discussed in Ref. \cite{Akhmedov}. Different from the
canonical seesaw model, the experimental testability of heavy
Majorana neutrinos is preserved in this scenario if the global
cancellation condition $M^{}_{\rm L} - M^{}_{\rm D} M^{-1}_{\rm R}
M^T_{\rm D} = {\bf 0}$ is satisfied \cite{type2}. Under this
condition, both heavy Majorana neutrinos and the doubly-charged
component of the Higgs triplet can be tested at the LHC via the
lepton-number-violating (LNV) processes \cite{CSXZ}.

In this paper, we propose a novel type-(I+II) seesaw model, in which
the contributions from heavy Majorana neutrinos to light neutrino
masses are vanishing $M^{}_{\rm D} M^{-1}_{\rm R} M^T_{\rm D} = {\bf
0}$. Thus the mass matrix of three known neutrinos is $M^{}_\nu =
M^{}_{\rm L}$, which is the same as in the pure type-II seesaw model
\cite{SS2}. However, the apparent difference of our scenario is that
the neutrino mixing matrix becomes non-unitary. Furthermore, the
heavy Majorana neutrinos can be discovered at the LHC as in the
testable canonical seesaw model. Recently, the collider signals of
three heavy Majorana neutrinos have been considered in Ref.
\cite{Pavel} in the type-I seesaw model, which extends the previous
works about only one heavy Majorana neutrino case \cite{Han}.
Another purpose of the present work is to consider the collider
signatures of more than one heavy Majorana neutrinos in a realistic
type-(I+II) seesaw model.

The remaining part of our paper is organized as follows. In Section
II, we propose an interesting type-(I+II) model with only two heavy
Majorana neutrinos, in which the structural cancellation condition
$M^{}_{\rm D} M^{-1}_{\rm R} M^T_{\rm D} = {\bf 0}$ is achieved by
imposing an ${\rm A^{}_4 \times Z^{}_2}$ flavor symmetry. As a
result of this elegant symmetry, the nearly tri-bimaximal neutrino
mixing can be obtained. Section III is devoted to calculating the
cross sections of the processes $pp \to l^\pm_\alpha N^{}_i \to
l^\pm_\alpha l^\pm_\beta j j$, where $N^{}_i$ (for $i=1, 2$) are the
heavy Majorana neutrinos. The interference effects of different
heavy Majorana neutrinos are emphasized, and the possible way to
distinguish between the same-sign dilepon signals from heavy
Majorana neutrinos and the doubly-charged Higgs bosons is briefly
discussed. Furthermore, the background from the SM have been taken
into account, and the kinematic cuts are imposed to efficiently
select the signal events. Finally, we give some concluding remarks
in Section IV.

\section{Testable Type-(I+II) Seesaw Models}

In order to generate tiny neutrino masses, we can extend the SM by
introducing three right-handed neutrinos and a triplet scalar. The
gauge-invariant Lagrangian relevant for lepton masses can be written
as
\begin{eqnarray}
-{\cal L}^{}_{\rm lepton} = \overline{\ell^{}_{\rm L}} Y^{}_l
E^{}_{\rm R} H + \overline{\ell^{}_{\rm L}} Y^{}_\nu N^{}_{\rm R}
\tilde{H} + \frac{1}{2} \overline{N^c_{\rm R}} M^{}_{\rm R}
N^{}_{\rm R} + \frac{1}{2} \overline{\ell^{}_{\rm L}} Y^{}_\Delta
\Delta i \sigma^{}_2 \ell^c_{\rm L} + {\rm h.c.} \; ,
\end{eqnarray}
where $\ell^{}_{\rm L}$ and $\tilde{H} \equiv i\sigma^{}_2 H^*$ are
respectively the lepton and Higgs doublets, $E^{}_{\rm R}$ and
$N^{}_{\rm R}$ are the right-handed charged-lepton and neutrino
singlets, $\Delta$ is the triplet scalar in the $2\times 2$ matrix
form. After the spontaneous gauge symmetry breaking, the lepton mass
terms turn out to be
\begin{equation}
-\mathcal{L}^{}_{\rm m} = \overline{e^{}_{\rm L}} M^{}_l E^{}_{\rm
R} + \frac{1}{2} \overline{(\nu^{}_{\rm L} ~~ N^c_{\rm R})}
\left(\matrix{M^{}_{\rm L} & M^{}_{\rm D} \cr M^T_{\rm D} &
M^{}_{\rm R}}\right) \left(\matrix{\nu^c_{\rm L} \cr N^{}_{\rm
R}}\right) + {\rm h.c.} \; ,
\end{equation}
where $M^{}_{\rm L} = Y^{}_\Delta v^{}_\Delta$ and $M^{}_{\rm D} =
Y^{}_\nu v/\sqrt{2}$ are the Majorana and Dirac neutrino mass terms
with $v$ and $v^{}_\Delta$ being the vacuum expectation values of
the doublet and triplet scalars, respectively. $M^{}_l$ and
$M^{}_{\rm R}$ are the charged-lepton and heavy right-handed
Majorana neutrino mass matrices. The total $6\times 6$ neutrino mass
matrix can be diagonalized by the following unitary transformation
\begin{eqnarray}
\left(\matrix{V & R \cr S & U}\right)^\dagger
\left(\matrix{M^{}_{\rm L} & M^{}_{\rm D} \cr M^T_{\rm D} &
M^{}_{\rm R}}\right) \left(\matrix{V & R \cr S & U}\right)^* =
\left(\matrix{\widehat{M}^{}_\nu & \mathbf{0} \cr \mathbf{0} &
\widehat{M}^{}_{\rm N}}\right) \; ,
\end{eqnarray}
where $\widehat{M}^{}_\nu = {\rm Diag}\{m^{}_1, m^{}_2, m^{}_3\}$
and $\widehat{M}^{}_{\rm N} = {\rm Diag}\{M^{}_1, M^{}_2, M^{}_3\}$
are the mass eigenvalues of light and heavy Majorana neutrinos,
respectively. In the leading-order approximation, the effective
neutrino mass matrix is determined by the seesaw formula
\begin{eqnarray}
M^{}_\nu = M^{}_{\rm L} - M^{}_{\rm D} M^{-1}_{\rm R} M^T_{\rm D} \;
.
\end{eqnarray}
Thus the smallness of light neutrino masses are attributed to the
heaviness of right-handed neutrinos and the smallness of $M^{}_{\rm
L}$. It is obvious that the above equation reduces to the canonical
seesaw formula $M^{}_\nu = - M^{}_{\rm D} M^{-1}_{\rm R} M^T_{\rm
D}$, if the triplet scalar is absent. In the basis where the mass
eigenstates of charged leptons coincide with their flavor
eigenstates, the leptonic charged-current interactions can be
expressed as
\begin{eqnarray}
-{\cal L}^{}_{\rm CC} = \frac{g}{\sqrt{2}} \overline{\left(\matrix{e
& \mu & \tau}\right)^{}_{\rm L}} \gamma^\mu \left[V
\left(\matrix{\hat{\nu}^{}_1 \cr \hat{\nu}^{}_2 \cr
\hat{\nu}^{}_3}\right)^{}_{\rm L} + R \left(\matrix{\hat{N}^{}_1 \cr
\hat{N}^{}_2 \cr \hat{N}^{}_3}\right)^{}_{\rm L}\right] W^-_\mu +
{\rm h.c.} \; ,
\end{eqnarray}
where $\hat{\nu}^{}_i$ and $\hat{N}^{}_i$ (for $i=1, 2, 3$) stand
for the mass eigenstates of three light and heavy Majorana
neutrinos, respectively. It is the charged-current interactions that
govern both the production and detection of heavy Majorana neutrinos
in the hadron collider experiments. From Eq. (5) and the unitarity
condition $VV^\dagger + RR^\dagger = {\bf 1}$, we can see that the
neutrino mixing matrix $V$ is non-unitary. Therefore, both the
detection of heavy Majorana neutrinos and leptonic unitarity
violation are determined by the couplings $R^{}_{\alpha i}$ (for
$\alpha = e, \mu, \tau$ and $i = 1, 2, 3$), which represent distinct
interaction strengths of heavy Majorana neutrinos with charged
leptons. In the conventional type-(I+II) seesaw model, both terms on
the right-hand side of Eq. (4) are comparable in magnitude and on
the same order of the masses of three light neutrinos. In this case,
the masses of heavy degrees of freedom are expected to be around the
scale of grand unified theories, i.e. $\Lambda^{}_{\rm GUT} =
10^{16}~{\rm GeV}$, so the conventional seesaw models can not be
tested experimentally. In order that the heavy Majorana neutrinos
can be produced and detected at the LHC, one should appeal to the
following scenarios:
\begin{itemize}
\item {\it Scenario A} with ${\cal O}(M^{}_{\rm L}) \ll {\cal
O}(M^{}_\nu)$ and ${\cal O}(M^{}_{\rm D} M^{-1}_{\rm R} M^T_{\rm D})
\sim {\cal O}(M^{}_\nu)$, but ${\cal O}(M^{}_{\rm R}) \sim {\cal
O}(1~{\rm TeV})$ and ${\cal O}(R) \sim {\cal O}(M^{}_{\rm D}
M^{-1}_{\rm R}) \lesssim 10^{-1}$. This is similar to the canonical
seesaw model, where the structural cancellation condition $M^{}_{\rm
D} M^{-1}_{\rm R} M^T_{\rm D} \approx {\bf 0}$ is required to render
heavy Majorana neutrinos testable \cite{Pilaftsis,Smirnov}.

\item {\it Scenario B} with ${\cal O}(M^{}_{\rm L}) \sim {\cal O}(M^{}_{\rm D}
M^{-1}_{\rm R} M^T_{\rm D})  \gg {\cal O}(M^{}_\nu)$, but ${\cal
O}(M^{}_{\rm L} - M^{}_{\rm D} M^{-1}_{\rm R} M^T_{\rm D}) \sim
{\cal O}(M^{}_\nu)$. This implies that the significant but
incomplete global cancellation exists between $M^{}_{\rm L}$ and
$M^{}_{\rm D} M^{-1}_{\rm R} M^T_{\rm D}$ \cite{type2}. In this
case, the collider signals at the LHC induced by heavy Majorana
neutrinos and the doubly-charged component of the triplet scalar are
correlated with each other \cite{CSXZ}.

\item {\it Scenario C} with ${\cal O}(M^{}_{\rm L}) \sim {\cal
O}(M^{}_\nu)$ and ${\cal O}(M^{}_{\rm D} M^{-1}_{\rm R} M^T_{\rm D})
\ll {\cal O}(M^{}_\nu)$. The latter condition is consistent with the
structural cancellation condition $M^{}_{\rm D} M^{-1}_{\rm R}
M^T_{\rm D} \approx {\bf 0}$, so both heavy Majorana neutrinos and
the triplet scalar can be discovered at the LHC. This interesting
scenario has also been discussed in Ref. \cite{Gu} in a very
different context.
\end{itemize}
It has been observed that {\it Scenario B} may suffer from the
problem of radiative instability and then serious fine-tunings
\cite{CSXZ}. This drawback can be avoided in {\it Scenario C}, if
the heavy Majorana neutrinos are nearly degenerate in mass
\cite{Pilaftsis,Smirnov}.

In the following, we shall concentrate on {\it Scenario C} and
propose an interesting model with several additional scalar fields
and two heavy Majorana neutrinos, which may serve as a
straightforward extension of the minimal type-(I+II) seesaw model
\cite{Minimal2,CSXZ}. It is worth mentioning that our discussions
can easily be made applicable to the case with three heavy Majorana
neutrinos, however, only two of them are sufficient for our purpose
\cite{GXZ}. To realize {\it Scenario C}, we impose the ${\rm A^{}_4
\times Z^{}_2}$ flavor symmetry on the generic Lagrangian of the
Type-(I+II) seesaw model. The assignments of relevant lepton and
scalar fields with respect to the symmetry group ${\rm SU(2)_{\rm
L}^{} \times U(1)_{\rm Y}^{}} \otimes {\rm A_4^{} \times Z_2^{}}$
are summarized as follows
\begin{eqnarray}
&& \ell^{}_{\rm L} \sim (2, -1) \otimes (\underline{3}, 1) \; ,
~~~~~~~~~
H \sim (2, 1) \otimes (\underline{3}, 1) \; , \nonumber \\
&& E^{}_{\rm R} \sim (1, 1) \otimes (\underline{1}, 1) \; , \;\;\;
~~~~~~~
\phi \sim (1, 0) \otimes (\underline{1}, -1) \; , \nonumber \\
&& E^\prime_{\rm R} \sim (1, -2) \otimes (\underline{1}^\prime, 1)
\; , ~~~~~~~ \Delta \sim (3, -2) \otimes (\underline{3}, 1) \; , \nonumber \\
&& E^{\prime \prime}_{\rm R} \sim (1, -2) \otimes
(\underline{1}^{\prime \prime}, 1) \; , ~~~~~~ \Sigma \sim (3, -2)
\otimes (\underline{1}, 1) \; , \nonumber \\
&& N^{}_{\rm R} \sim (1, 0) \otimes (\underline{1^{\prime\prime}},
1) \; , ~~~~~~~~~ N^\prime_{\rm R} \sim (1, 0) \otimes
(\underline{1^{\prime}}, -1) \;,
\end{eqnarray}
where three scalar doublets $H^{}_i$ (for $i = 1, 2, 3$), four
scalar triplets $\Sigma$ and $\Delta^{}_i$ (for $i = 1, 2, 3$), and one
scalar singlet $\phi$ have been introduced. The gauge- and ${\rm
A_4^{} \times Z_2^{}}$-invariant Lagrangian responsible for lepton
masses turns out to be
\begin{eqnarray}
-{\cal L}_{\rm lepton}^{\prime} &=& y_{1}^{l} (\overline{\ell_{\rm L
}^{}}H )_{\underline{1}}^{} E_{\rm R}^{} +
y_{2}^{l}(\overline{\ell_{\rm L}^{}}H )_{\underline{1^\prime}}^{}
E_{\rm R}^{\prime\prime}+ y_{3}^{l}(\overline{\ell_{\rm L}^{}}H
)_{\underline{1^{\prime\prime}}}^{} E_{\rm R}^{\prime}+y_{\nu}^{}
(\overline{\ell_{\rm L }^{}}\tilde{H} )_{\underline{1^\prime}}^{}
N_{\rm R}^{} \nonumber \\
&& + {1\over 2} y_\Sigma^{} \overline{\ell_{\rm L}^{}} \Sigma
i\sigma_2^{} \ell_{\rm L}^{\rm c}+{1\over 2} y_{\Delta}^{}
\overline{\ell_{\rm L}^{}} \Delta i\sigma_2^{} \ell_{\rm L}^{\rm c}
+ y_N^{}\overline{N_{\rm R}^c} N_{\rm R}^\prime \phi  +{\rm h.c.}
\;.
\end{eqnarray}
Given the irreducible representations and multiplication rules of
the ${\rm A^{}_4}$ group in \cite{He,A4}, one can immediately verify
the ${\rm A_4^{} \times Z_2^{}}$-invariance of ${\cal L}^\prime_{\rm
lepton}$. After the spontaneous gauge symmetry breaking, the mass
matrix of charged leptons is
\begin{eqnarray}
M_l^{}= \left(\matrix{ y_{1}^{l} v_1^{} & y_{3}^{l} v_1^{} &
y_{2}^{l} v_1^{} \cr  y_{1}^{l} v_2^{} & y_{3}^{l} v_2^{} \omega &
y_{2}^{l} v_2^{} \omega^2  \cr y_{1}^{l} v_3^{} & y_{3}^{l} v_3^{}
\omega^2 & y_{2}^{l} v_3^{} \omega  }\right) \; ;
\end{eqnarray}
and the neutrino mass matrices are
\begin{eqnarray}
M_{\rm L}^{} = \left(\matrix{ y^{}_\Sigma v^{}_\Sigma & y^{}_\Delta
u^{}_3 &  y^{}_\Delta u^{}_2 \cr y^{}_\Delta u^{}_3 & y_\Sigma^{}
v^{}_\Sigma &  y^{}_\Delta u_{1}^{} \cr y^{}_\Delta u_{2}^{} &
y^{}_\Delta u_{1}^{} & y_\Sigma^{} v_\Sigma^{}} \right) \; ,~~~
M_{\rm D}^{} = \left ( \matrix{y_\nu^{} v_1^{} & 0 \cr y_\nu^{}
v_2^{} \omega^2 & 0 \cr y_\nu^{} v_3^{} \omega & 0 }\right) \; ,~~~
M_{\rm R}^{} = \left( \matrix{ 0 & y_N^{} v_\phi\cr y_N^{} v_\phi &
0 } \right) \; ,~~
\end{eqnarray}
where the vacuum expectation values are taken to be $v_i^{} =
\langle H_i^{} \rangle$ (for $i = 1, 2, 3$), $u^{}_i = \langle
\Delta_i^{} \rangle$ (for $i = 1, 2, 3$), $v_\Sigma^{}=\langle
\Sigma \rangle$ and $v_\phi =\langle \phi \rangle$, while $\omega =
\exp(2i\pi/3)$ is the cubic root of $+1$. Provided the textures of
$M^{}_{\rm D}$ and $M^{}_{\rm R}$ in Eq. (9), we can see that
$M_{\rm D}^{} M_{\rm R}^{-1} M_{\rm D}^T = {\bf 0}$ holds exactly.
As a consequence, the non-zero neutrino masses mainly arises from
the Type-II seesaw mechanism $M^{}_\nu = M^{}_{\rm L}$. As implied
by Eq. (8) with the assumption that $v_1^{}=v_2^{}=v_3^{} \equiv
v^{}_H$, the charged-lepton mass matrix can be written as $M_l^{} =
U_l^{} \cdot {\rm Diag} \{\sqrt{3} y_{1}^{l} v^{}_H, \sqrt{3}
y_{3}^{l} v^{}_H, \sqrt{3} y_{2}^{l} v^{}_H \} $, which is simply
diagonalized by the unitary matrix
\begin{eqnarray}
U_l^{}= {1 \over \sqrt{3}}\left( \matrix{1&1&1 \cr 1 & \omega &
\omega^2 \cr 1 & \omega^2 & \omega} \right) \; ;
\end{eqnarray}
Therefore, the masses of charged leptons are identified as $m^{}_e =
\sqrt{3} y_{1}^{l} v^{}_H$, $m^{}_\mu = \sqrt{3} y_{3}^{l} v^{}_H$
and $m^{}_\tau = \sqrt{3} y_{2}^{l} v^{}_H$. Setting $u^{}_1 =
u^{}_3 = 0$ and $u^{}_2 \neq 0$, we can obtain the neutrino mass
matrix
\begin{eqnarray}
M_\nu^{} = \left( \matrix{ y_\Sigma^{} v_\Sigma^{} & 0 & y_\Delta^{}
u_{2}^{} \cr 0 & y_\Sigma^{} v_\Sigma^{} & 0 \cr  y_\Delta^{}
u_{2}^{} & 0 &y_\Sigma^{} v_\Sigma^{} } \right) \; ,
\end{eqnarray}
which can be diagonalized by the $\pi/4$-rotation in the $1$-$3$
plane, namely
\begin{eqnarray}
U_\nu^{}= {1\over \sqrt{2}}\left( \matrix{1 & 0 & -1 \cr 0 &
\sqrt{2} & 0 \cr 1 & 0& 1}\right) \; .
\end{eqnarray}
The mass eigenvalues of three known neutrinos are then given by
$m_1^{}= |y_\Sigma^{} v_\Sigma^{} + y_\Delta^{} u_{2}^{}|$, $m_2^{}
= |y_\Sigma^{} v_\Sigma^{}|$ and $m_3^{}= |y_\Sigma^{}
v_\Sigma^{}-y_\Delta^{} u_{2}^{}|$, which can fit the observed
values of two neutrino mass-squared differences $\Delta m^2_{21}
\equiv m^2_2 - m^2_1 = 8.0 \times 10^{-5}~{\rm eV}^2$ and $|\Delta
m^2_{32}| \equiv |m^2_3 - m^2_2| = 2.5 \times 10^{-3}~{\rm eV}^2$
\cite{Review}. In the leading-order approximation, the non-unitary
lepton flavor mixing matrix $V$ is just the unitary matrix $V^{}_0$,
which arises from the mismatch between the diagonalizations of
$M_l^{}$ and $ M_\nu^{}$. To be more explicit,
\begin{eqnarray}
V \approx V^{}_0 = U_l^{\dagger} U_\nu^{} = \left (
\matrix{\displaystyle \frac{2}{\sqrt{6}} &
\displaystyle\frac{1}{\sqrt{3}}& 0 \cr \displaystyle
-\frac{1}{\sqrt{6}} \omega^2 & \displaystyle \frac{1}{\sqrt{3}}
\omega^2 & \displaystyle -\frac{1}{\sqrt{2}} e^{-i \pi/6} \cr
\displaystyle - \frac{1}{\sqrt{6}} \omega & \displaystyle
\frac{1}{\sqrt{3}} \omega & \displaystyle - \frac{1}{\sqrt{2}}
e^{+i\pi / 6} }\right ) \; ,
\end{eqnarray}
which is equivalent to the tri-bimaximal mixing pattern \cite{Tribi}
strongly favored by current neutrino oscillation experiments. From
Eq. (9), we can observe that two heavy Majorana neutrinos are
degenerate in mass. Two remarks are in order: (1) Because of
$M^{}_{\rm D} M^{-1}_{\rm R} M^T_{\rm D} = {\bf 0}$ and $M^{}_1 =
M^{}_2$, the one-loop radiative corrections to light neutrino masses
are extremely small and can be neglected \cite{Pilaftsis}; (2) The
slight breaking of the ${\rm A^{}_4 \times Z^{}_2}$ symmetry may
lead to a tiny mass split of heavy Majorana neutrinos, and then the
resonant leptogenesis mechanism can be implemented to account for
the baryon number asymmetry in the Universe \cite{Pilaftsis2}.

To be more accurate, we can get the masses of three light and two
heavy Majorana neutrinos after diagonalizing the total $5\times 5$
neutrino mass matrix by a unitary transformation. The full
parametrization of the corresponding $5\times 5$ unitary matrix will
involve 10 rotation angles $\theta^{}_{ij}$ and 10 phase angles
$\delta^{}_{ij}$ (for $i,j = 1, 2, \cdots, 5$ and $i<j$). Following
Ref. \cite{Xing}, we may adopt the standard parametrization of the
neutrino mixing matrix $V = A V^{}_0$, where $V^{}_0$ is a $3\times
3$ unitary matrix
\begin{eqnarray}
V^{}_0 = \left(\matrix{c^{}_{12} c^{}_{13} & \hat{s}^*_{12}
c^{}_{13} & \hat{s}^*_{13} \cr -\hat{s}^{}_{12}c^{}_{23} - c^{}_{12}
\hat{s}^{}_{13} \hat{s}^*_{23} & c^{}_{12} c^{}_{23} -
\hat{s}^*_{12} \hat{s}^{}_{13} \hat{s}^*_{23} & c^{}_{13}
\hat{s}^*_{23} \cr \hat{s}^{}_{12} \hat{s}^{}_{23} - c^{}_{12}
\hat{s}^{}_{13} c^{}_{23} & -c^{}_{12} \hat{s}^{}_{23} -
\hat{s}^*_{12} \hat{s}^{}_{13} c^{}_{23} & c^{}_{13}
c^{}_{23}}\right) \;
\end{eqnarray}
and $A$ is nearly an identity matrix
\begin{eqnarray}
A = {\bf 1} - \sum^5_{j=4} \left(\matrix{s^2_{1j} & 0 & 0 \cr
\hat{s}^{}_{1j} \hat{s}^*_{2j} & s^2_{2j} & 0 \cr \hat{s}^{}_{1j}
\hat{s}^*_{3j} & \hat{s}^{}_{2j} \hat{s}^*_{3j} & s^2_{3j}}\right) +
{\cal O}(s^4_{ij}) \; ,
\end{eqnarray}
with $c^{}_{ij} \equiv \cos \theta^{}_{ij}$, $s^{}_{ij} \equiv \sin
\theta^{}_{ij}$ and $\hat{s}^{}_{ij} \equiv e^{i\delta^{}_{ij}}
s^{}_{ij}$ (for $1\leq i < j \leq 5$). It is worthwhile to note that
the deviation of $A$ from the identity matrix measures the unitary
violation of neutrino mixing matrix, which is constrained to be
below the percent level \cite{Antusch}. Therefore, it is an
excellent approximation to neglect the higher-order terms ${\cal
O}(s^4_{ij})$ in Eq. (15). Meanwhile, the parametrization of $R$ can
be taken as
\begin{equation}
R = {\bf 0} + \left(\matrix{\hat{s}^*_{14} & \hat{s}^*_{15} \cr
\hat{s}^*_{24} & \hat{s}^*_{25} \cr \hat{s}^*_{34} &
\hat{s}^*_{35}}\right) + {\cal O}(s^3_{ij}) \; .
\label{eq:parameter}
\end{equation}
It has been stressed in Ref. \cite{Xing} that the charged-current
interactions of light and heavy Majorana neutrinos are correlated,
which is obvious from Eq. (15) and Eq. (16). Furthermore, the extra
CP-violating phases come into the non-unitary neutrino matrix $V$
via the matrix $A$. Thus novel CP-violating effects in the
medium-baseline $\nu^{}_\mu \to \nu^{}_\tau$ and
$\overline{\nu}^{}_\mu \to \overline{\nu}^{}_\tau$ oscillations may
show up and provide a promising signature of the unitarity violation
of $V$, which could be measured at a neutrino factory
\cite{Xing,CP}.

\section{Collider Signals of Heavy Majorana Neutrinos}

As pointed out in Ref. \cite{Smirnov}, the generation of neutrino
masses and the collider signals of heavy Majorana neutrinos
essentially decouple in the realistic seesaw model. Therefore, a
phenomenological approach to consider collider signals of heavy
Majorana neutrinos is to take the matrix elements $R^{}_{\alpha i}$
and the heavy Majorana neutrino masses $M^{}_{i}$ (for $i=1, 2, 3$)
as independent parameters, which should be consistent with both
low-energy and current collider experiments. This phenomenological
approach has been widely used in the literature \cite{Han}, however,
only for the one heavy Majorana neutrino case. We now generalize
previous works and include one more heavy Majorana neutrino, which
should be present in the realistic type-I seesaw model \cite{GXZ}.

Given the charged-current interactions in Eq. (5), the relevant
process reads
\begin{equation}
q(p_1) + \bar{q}^\prime (p_2) \to l^{\pm}_\alpha (p_3) N^{}_i (p)\to
l^\pm_\alpha (p_3)+ l^\pm_\beta (p_4) + q_f (p_5)+ \bar{q}_f^\prime
(p_6) \; , \label{eq:process}
\end{equation}
where $\alpha, \beta = e, \mu, \tau$, $i=1,2,3$ and $p^{}_1$,
$p^{}_2$ {\it etc.} represent the four-momentum of the corresponding
particles. Heavy Majorana neutrinos can be produced on-shell in this
channel, thus it is safe to neglect the contributions from the
$t$-channel diagrams. For simplicity, we consider the typical
example with two heavy Majorana neutrinos $N^{}_1$ and $N^{}_2$,
while the general situation with more heavy Majorana neutrinos can
be analyzed in a similar way. The squared matrix elements for the
process in Eq. (\ref{eq:process}) can be obtained as follows
\begin{eqnarray}
 \overline{|{\cal M}^{}_{N}|^2} &=&  g^8 \left(2 -
\delta^{}_{\alpha \beta}\right) \left|D^{}_W (\hat{s}) D^{}_W
(q^2)\right|^2 \left(p^{}_2 \cdot p^{}_6\right) \left\{ {M^2_1
\left|R^{}_{\alpha 1} R^{}_{\beta 1}\right|^2} {\cal F}^{}_1 \right.
\nonumber \\  && ~~~ + \left. M^2_2 \left|R^{}_{\alpha 2}
R^{}_{\beta 2}\right|^2 {\cal F}^{}_2 + M^{}_1 M^{}_2
\left|R^{}_{\alpha 1} R^{}_{\alpha 2} R^{}_{\beta 1} R^{}_{\beta
2}\right| {\rm Re}\left[{\cal G}^* e^{i\delta}\right]\right\} \; ,
\label{eq:amplitude}
\end{eqnarray}
where $\hat{s} \equiv (p^{}_1 +p^{}_2)^2$, $q \equiv p^{}_5 + p^{}_6$,
$\delta \equiv (\delta^{}_{\alpha 1} - \delta^{}_{\alpha 2}) +
(\delta^{}_{\beta 1} - \delta^{}_{\beta 2})$, and $R_{\alpha i}$
($\alpha =e, \mu, \tau$ and $i = 1, 2$) are the mixing matrix elements
as indicated in Eq. (\ref{eq:parameter}). The explicit expressions of
relevant functions $D^{}_W$, ${\cal F}^{}_i$ and ${\cal G}$ are
collected in Appendix A.

At the hardon collider, the total cross section for the process in
Eq. (\ref{eq:process}) can be expressed as follows
\begin{equation}
\sigma=\sum_{a,b} \int {\rm d}x^{}_1 {\rm d}x^{}_2 F^{}_{a/p}(x^{}_1,Q^2)
\cdot F^{}_{b/p}(x^{}_2,Q^2) \cdot \hat{\sigma}(ab \to l^\pm_{\alpha}
l^\pm_{\beta} q_f \bar{q}^\prime_f) \; ,
\label{eq:crosssection}
\end{equation}
where $F^{}_{a,b/p}$ denote the parton distribution functions for
the proton, $x^{}_{1,2}$ the energy fractions of the partons $a$ and
$b$, $Q$ the factorization scale, and $\hat{\sigma}$ the partonic
cross section. In our calculations, we consider the reactions at the
LHC ($\sqrt{S} = 14~{\rm TeV}$) and set $|R^{}_{ei}|^2=1.25 \times
10^{-7}$, $|R^{}_{\mu i}|^2 = |R^{}_{\tau i}|^2 = 5 \times 10^{-3}$
(for $i=1, 2$) as well as $\delta = 0$. The factorizaton scale is
taken to be $Q^2 = \hat{s}$.  Since the detection of charged leptons
$\mu^\pm$ is most efficient at the LHC, it is reasonable to explore
the $pp \to \mu^\pm \mu^\pm j j$ processes.

\begin{figure}
\centering ~~
\includegraphics[width=0.42\textwidth]{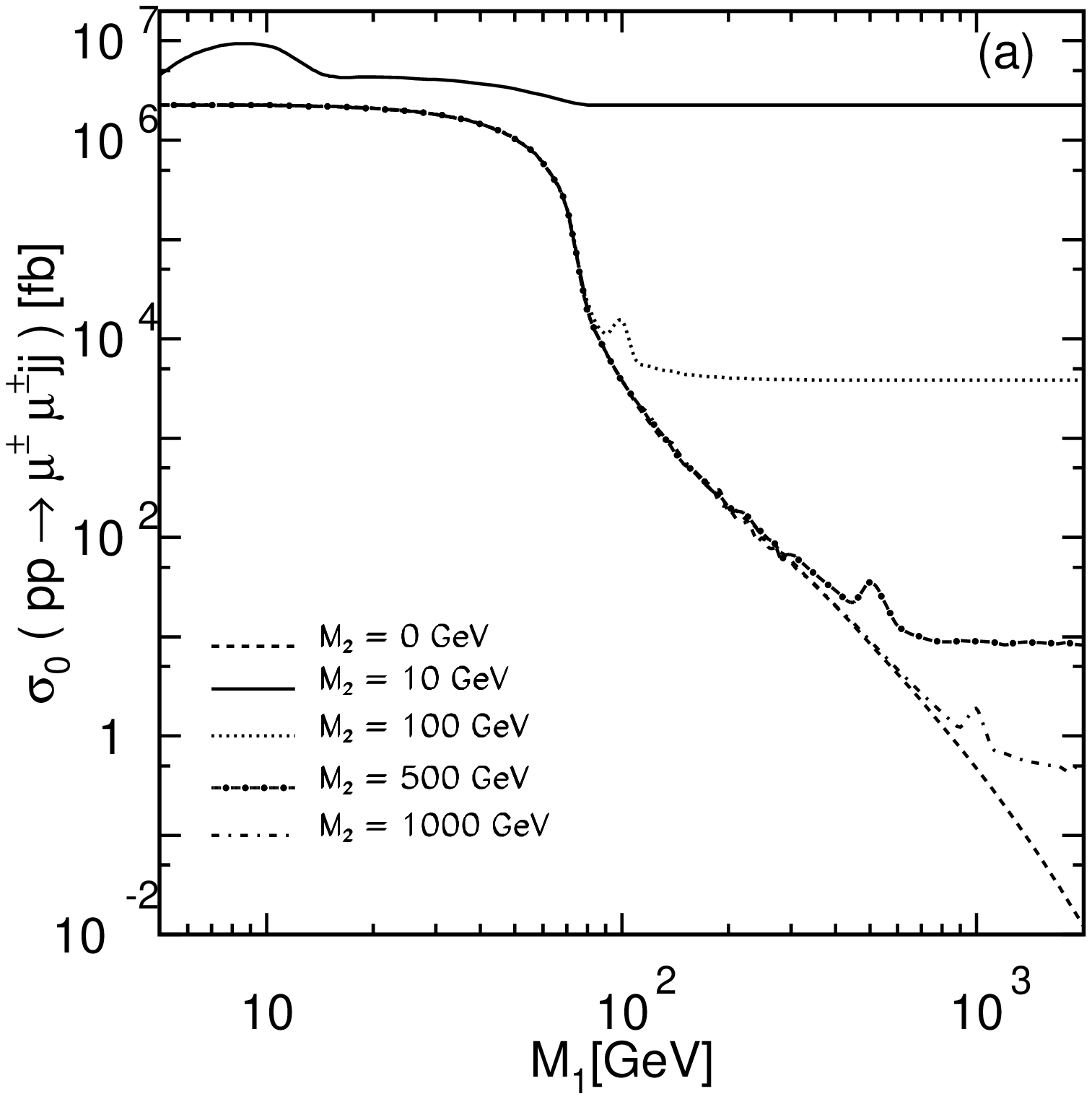} ~~~~~
\includegraphics[width=0.42\textwidth]{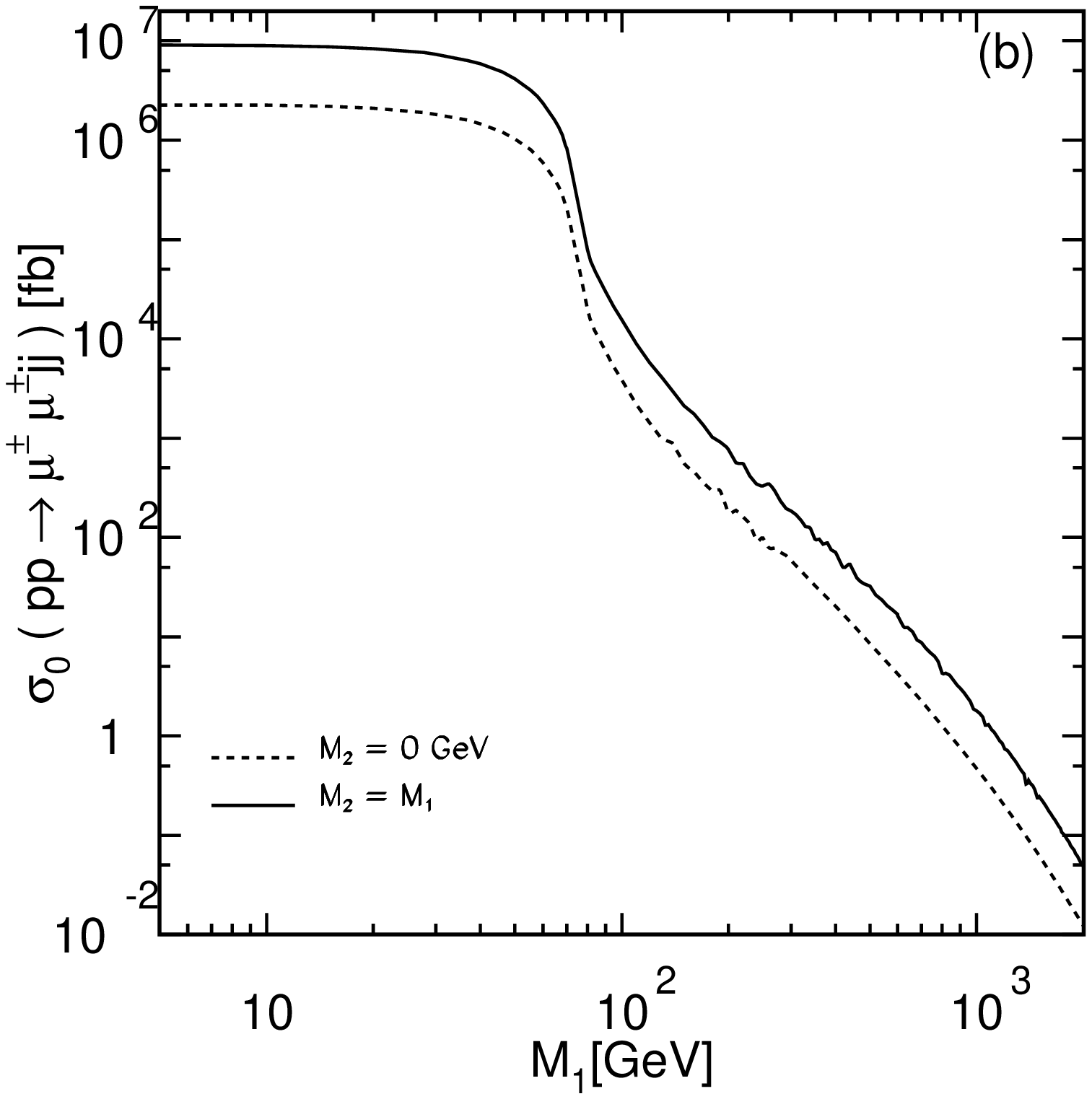}
\caption{(a) The cross sections for the same-sign dilepton events
$pp \to \mu^{\pm} \mu^{\pm} j j$ with $M^{}_2 = 10~{\rm GeV}$ (solid
line), $100~{\rm GeV}$ (dotted line), $500~{\rm GeV}$ (dotted-solid
line) and  $1~{\rm TeV}$ (dotted-dashed line) versus $M^{}_1$
varying from $5~{\rm GeV}$ to $2~{\rm TeV}$; (b) The solid line
corresponds to the case of $M^{}_1 = M^{}_2$. In both panels, the
dashed line represents for the reduced cross section with only one
heavy Majorana neutrino $N^{}_1$.} \label{fig:1}
\end{figure}

If only one heavy Majorana neutrino is taken into account, the
corresponding cross section for the processes can be decomposed as
\begin{equation}
\sigma (pp \to l^{\pm}_{\alpha} l^{\pm}_{\beta} j j) \approx (2-
\delta^{}_{\alpha \beta}) S^{}_{\alpha \beta} \sigma^{}_0 \; ,
\label{eq:reduction}
\end{equation}
with $S^{}_{\alpha \beta} \equiv \left|R^{}_{\alpha 1} R^{}_{\beta
1}\right|^2 / \sum_\gamma \left|R^{}_{\gamma 1}\right|^2$ and
$\sigma^{}_0$ the reduced cross section. Our results for $\sigma_0$
with only one Majorana neutrino roughly agree with those in Ref.
\cite{Han}. For comparison, the same definition for $\sigma_0$ is
adopted to calculate the reduced cross section in the case with
$N^{}_1$ and $N^{}_2$. In FIG. \ref{fig:1}(a), we fix the masses
$M^{}_2 = 10~{\rm GeV}$, $100~{\rm GeV}$, $500~{\rm GeV}$ and
$1~{\rm TeV}$, and change $M^{}_1$ continuously from 5 GeV to 2 TeV.
The resonant enhancement appears where the heavy Majorana neutrino
masses are equal $M^{}_2 = M^{}_1$. In the region with $M^{}_1 \gg
M^{}_2$, the cross section receives the dominant contribution from
$N^{}_2$ and thus is almost independent of the mass $M^{}_1$. We
also investigate the interesting case with two degenerate heavy
Majorana neutrinos, i.e. $M^{}_2 = M^{}_1$. In FIG. \ref{fig:1}(b),
it is obvious that the reduced cross section in the degenerate case
(solid line) is precisely four times of that with only one heavy
Majorana neutrino (dashed line) for the phase difference $\delta =
0$.

The key feature of our signal events is the effective reconstruction
of the two heavy Majorana neutrino masses from the final state
charged leptons ($l^{}_1$ and $l^{}_2$) and jets. Since the final
leptons are indistinguishable, it is helpful to define the
differential distribution ${\rm d} \sigma / {\rm d} M_{ljj} \equiv
({\rm d} \sigma / {\rm d} M_{l_{1}jj} + {\rm d} \sigma / {\rm d}
M_{l_{2}jj}) / 2$, where the invariant masses $M^{}_{l^{}_i jj}$
(for $i = 1, 2$) are constructed from the momenta of related
charged-leptons and those of the two jets. In FIG. 2, we show the
invariant mass distributions in two different cases: (a) $M^{}_1 =
60~{\rm GeV}$ and $M^{}_2 = 500~{\rm GeV}$; (b) $M^{}_1 = 100~{\rm
GeV}$ and $M^{}_2 = 115~{\rm GeV}$. It seems from the distribution
shape and peak positions that our approach to the reconstruction of
heavy Majorana neutrino masses is effective.
\begin{figure}
\centering ~~
\includegraphics[width=0.42\textwidth]{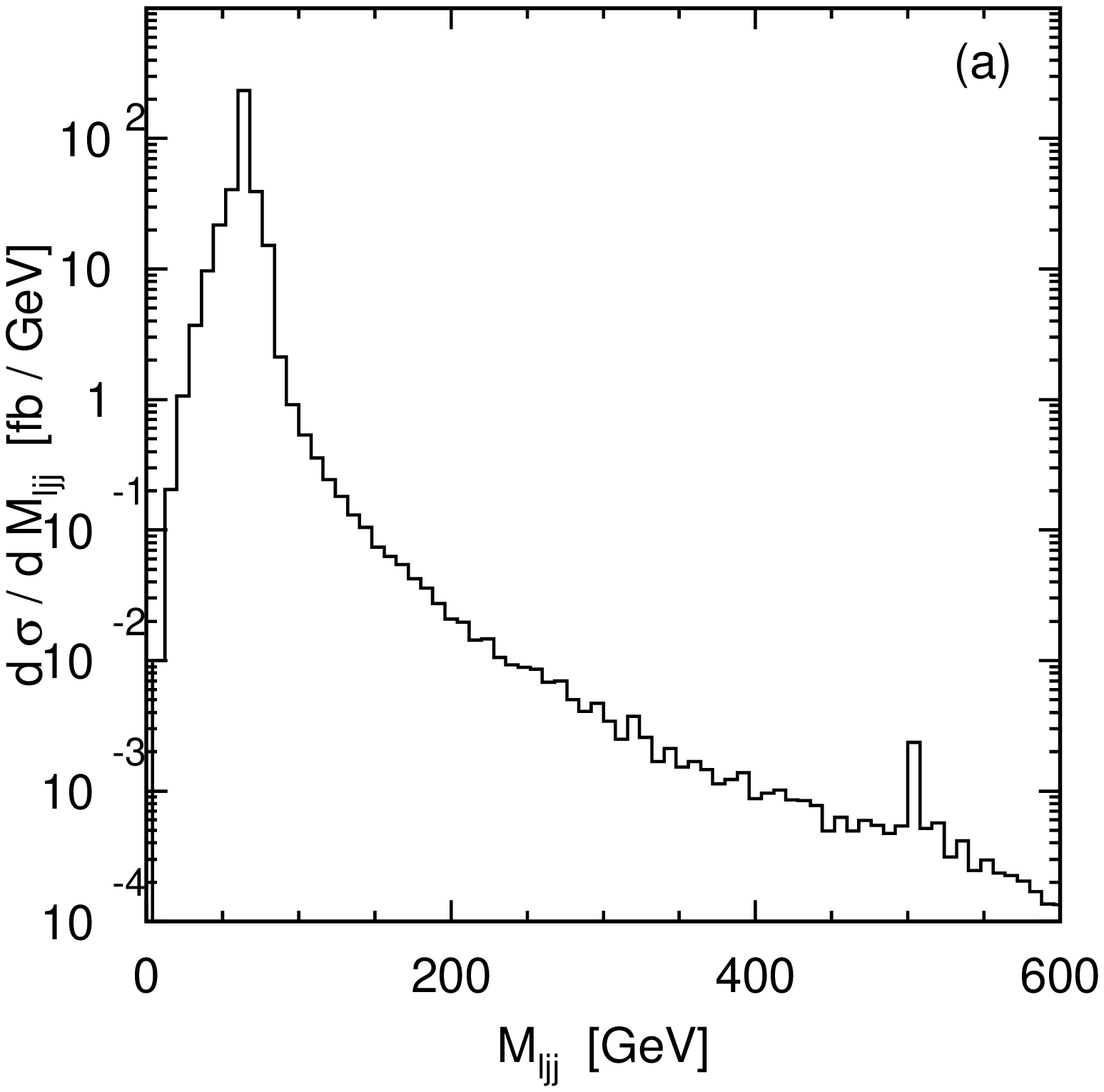} ~~~~~~
\includegraphics[width=0.42\textwidth]{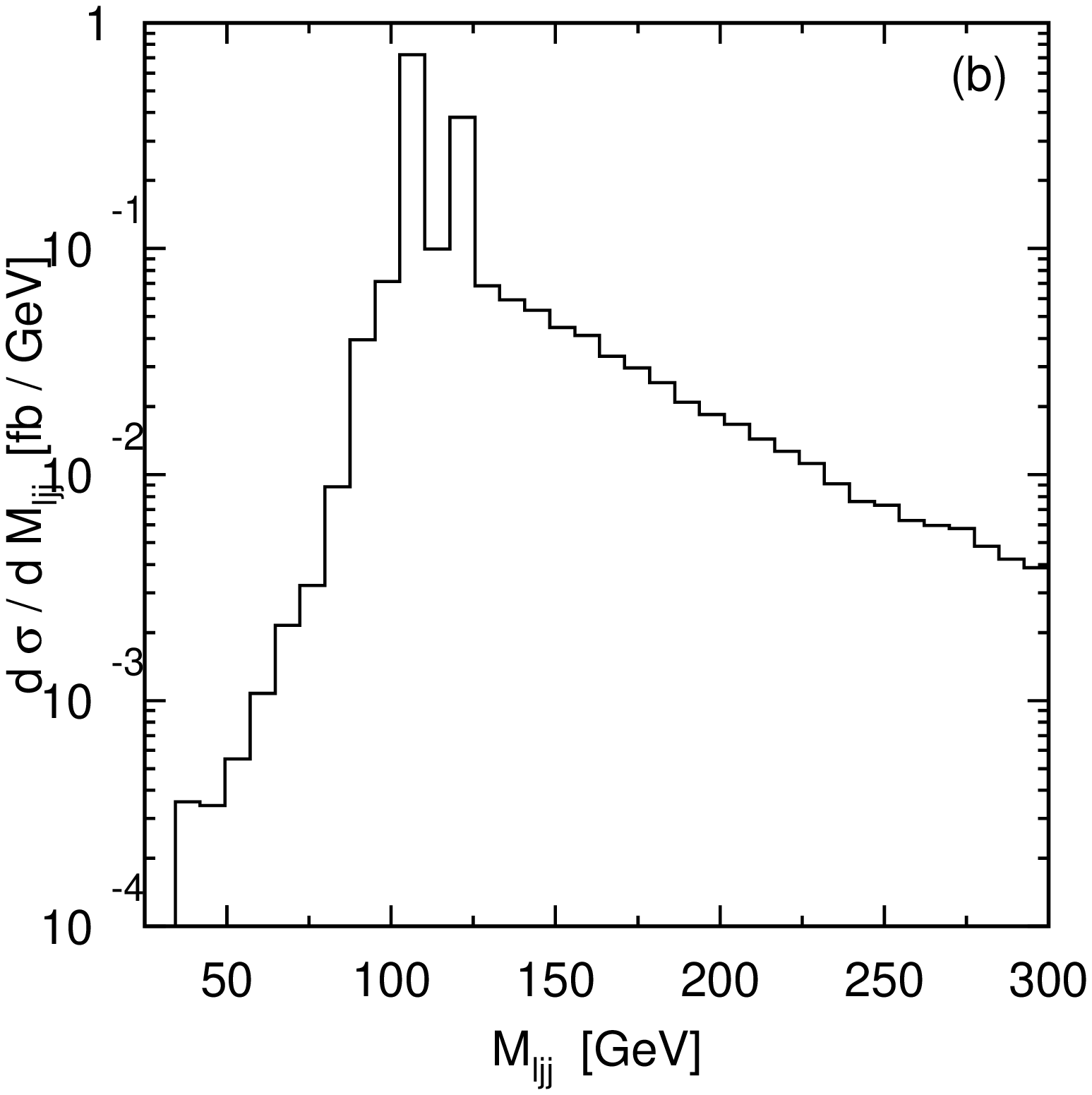} \vspace{0cm}
\caption{The invariant mass distribution of charged leptons and jets
with (a) $M^{}_1 = 60~{\rm GeV}, M^{}_2 = 500~{\rm GeV}$ and (b)
$M^{}_1 = 100~{\rm GeV}, M^{}_2 = 115~{\rm GeV}$.} \label{fig:2}
\end{figure}

To distinguish between the same-sign dilepton signals from heavy
Majorana neutrinos and those from the doubly-charged Higgs bosons
\cite{CSXZ}, we compute the differential distribution ${\rm
d}\sigma/{\rm d} \cos \theta^{}_{\mu \mu}$ and ${\rm d}\sigma/{\rm
d} M^{}_{\mu \mu}$ of the process in Eq. (17), where $\theta^{}_{\mu
\mu}$ is the angle between the final two leptons and $M_{\mu \mu}$
the invariant mass of them. The corresponding results are depicted
in FIG. 3(a) and FIG. 3(b) with $M^{}_1 (M^{}_2) = 60~(500)~{\rm
GeV}$ and $M^{}_1 (M^{}_2)= 100~(115)~{\rm GeV}$. In FIG. 3(a), it
is shown that the differential cross section decreases with
increasing $\cos \theta^{}_{\mu \mu}$. While for doubly-charged
Higgs bosons, the same-sign dileptons are from the decays of a
single scalar particle $H^{\pm \pm} \to \mu^\pm \mu^\pm$
\cite{Raidal,Han2,Chun,Han3}, the corresponding differential cross
section is independent of the $\theta^{}_{\mu \mu}$ angle. The
scalar particle decay processes also guarantee a peak around the
$H^{\pm \pm}$ mass ( $>$ 136 GeV \cite{Tevatron}) in the invariant
mass reconstruction of $M_{\mu \mu}$,  while for heavy Majorana
neutrinos there is no such signal as shown in FIG. 3(b). These two
distributions can serve as an excellent discriminator between the
same-sign dilepton siganls from heavy Majorana neutrinos and those
from the doubly-charged Higgs bosons.
\begin{figure}
\centering
\includegraphics[width=0.42\textwidth]{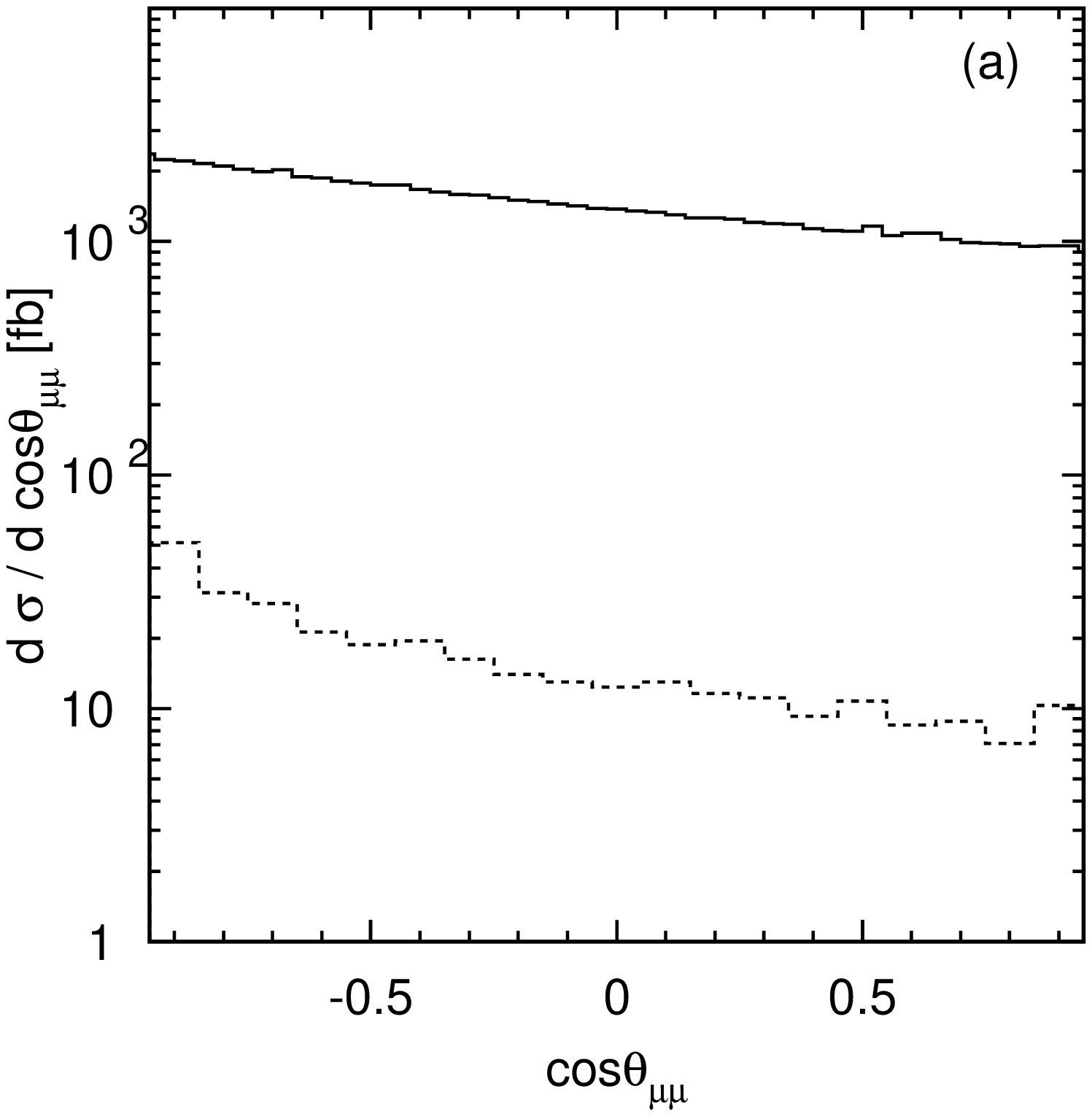} ~~~~~~
\includegraphics[width=0.43\textwidth]{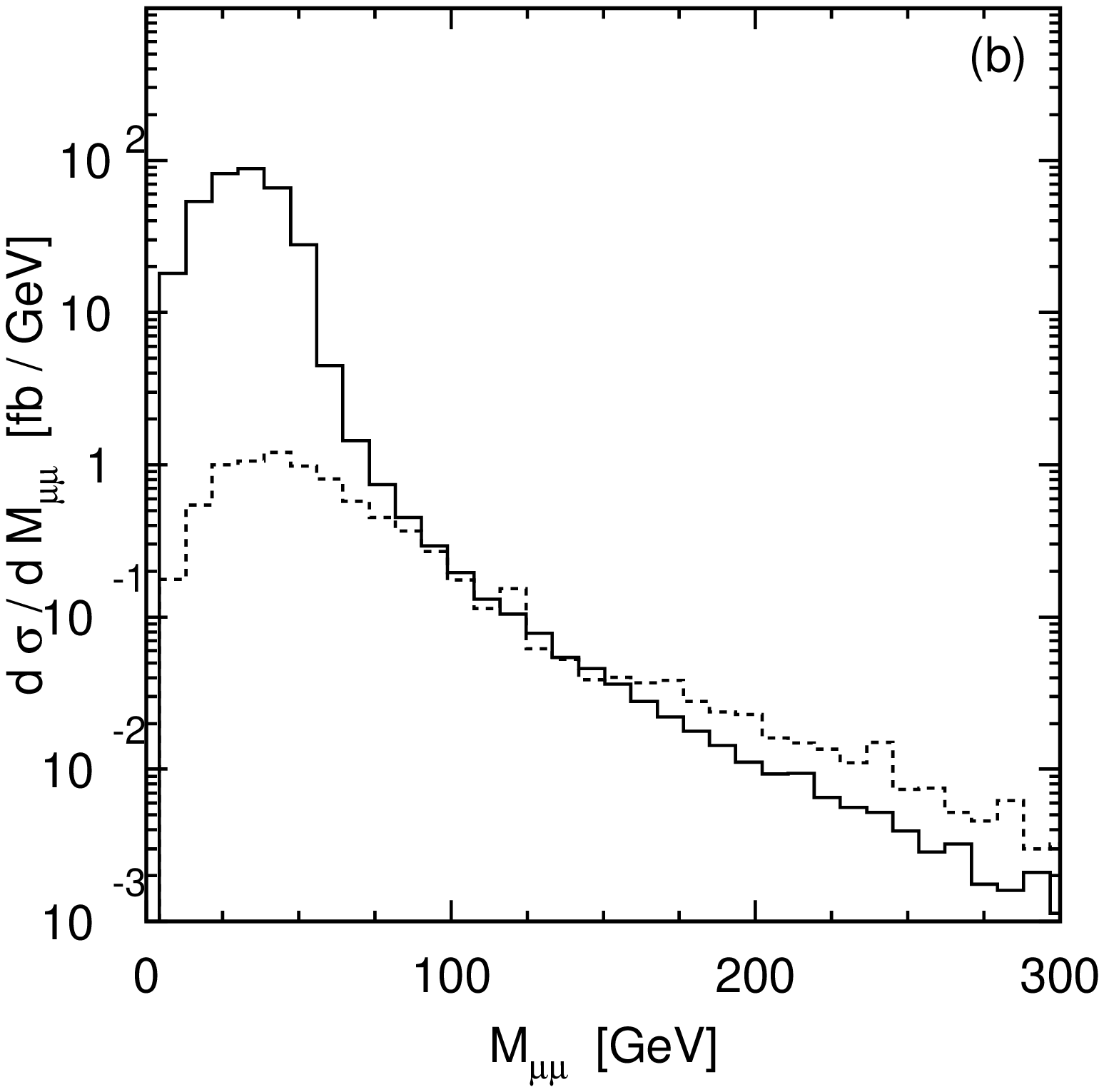}
\caption{(a) The angular distribution and (b) the invariant mass
distribution of the same-sign dilepton pair with $M^{}_1 = 60~{\rm
GeV}$, $M^{}_2 = 500~{\rm GeV}$ (solid line) and $M^{}_1 = 100~{\rm
GeV}$, $M^{}_2 = 115~{\rm GeV}$ (dashed line).} \label{fig:3}
\end{figure}

Now we turn to a brief discussion of the SM background and the
kinematic cuts used to reduce it. A salient feature of the process
in Eq. (17) is the same-sign dilepton with no missing energy in the
final states. However, due to the uncertainties in the measurement
of the jet energy and electromagnetic energy of charged leptons, the
missing transverse energy $\slashed{E}^{}_{\rm T}$ may appear. To
simulate the detector effects on the energy-momentum measurements,
we smear the charged-lepton (i.e., electron and muon) and jet
energies with a Gaussian distribution as follows
\begin{equation}
\frac{\Delta{E}}{E}=\frac{a}{\sqrt{E/\rm{GeV}}}\oplus{b} \; ,
\end{equation}
where $a^{}_l = 5\%$, $b^{}_l=0.55\%$ for charged leptons and
$a^{}_j = 100 \%, b^{}_j = 5\%$ for jets. The smearing simulation
shows that $\slashed{E}_{\rm T}$ cannot be neglected in the case of
$M^{}_1 = 60~\rm{GeV}$, $M^{}_2 = 500~\rm{GeV}$. Therefore, we
demand that there is no significant missing transverse energy
\begin{equation}
\slashed{E}_{\rm T} < 25~\rm{GeV} \; .
\end{equation}
Furthermore, we adopt the following basic cuts on charged-leptons
and jets
\begin{itemize}
\item $p^l_{\rm T} > 10~{\rm GeV}$ and $|\eta^{l}|<2.5$;
\item $p^j_{\rm T} > 20~{\rm GeV}$ and $|\eta^{j}|<2.5$;
\item $\Delta R_{lj} > 0.4 $,
\end{itemize}
where $p^{l,j}_{\rm T}$ stands respectively for the transverse
momentum of the charged lepton and jet, $\eta^{l,j}$ the
pseudo-rapidity, and $\Delta R^{}_{lj}=\sqrt{\Delta \eta^2 + \Delta \phi^2}$ the minimal isolation between any two of the final state leptons and jets.
At the LHC, the SM contribution to the like-sign dilepton events is
pretty small. The leading background comes from the top-quark pair production
and its cascade decays via the following chain
\begin{equation}
t \rightarrow W^+ b \rightarrow l^+_\alpha \nu_\alpha b \; , ~~~
\bar{t} \rightarrow W^- \bar{b} \rightarrow W^- \bar{c} \nu_\beta
l^+_\beta \; ,
\end{equation}
The signal and background cross
sections, as well as the efficiency of cuts, are given in TABLE I.
We see that the background events from $t\bar{t}$ is essentially
eliminated by the selective cut of missing transverse energy. In
addition, there are two other SM background processes coming from
like-sign $W$ boson production. First, the triple gauge-boson
production
\begin{equation}
pp\rightarrow W^{\pm}W^{\pm}W^{\mp}\rightarrow l^{\pm}l^{\pm}\nu \nu
jj \; ,
\end{equation}
leads to the irreducible background with two like-sign leptons plus
jets.  Second, the same final states can be produced via the process
\begin{equation}
pp\rightarrow W^{\pm}W^{\pm}jj\rightarrow l^{\pm}l^{\pm}\nu \nu jj
\; ,
\end{equation}
where the two jets may come either from QCD scattering or from the
gauge-boson fusion processes. In our calculations of the background
cross sections, we adopt the same couplings and conventions in
\cite{Madevent}. After imposing the cuts, we have found that these
backgrounds are extremely small, which have been listed in the last
two columns in TABLE I.
\begin{table}[t]
\begin{center}
\caption{The signal and background cross sections of $pp \to
\mu^{\pm}\mu^{\pm}jj$ at the LHC. In the calculation of the signal
cross section, we have taken $M^{}_1 = 60~{\rm GeV}$, $M^{}_2 = 500
~\rm{GeV}$ and $|R^{}_{\mu i}|^{2} = |R^{}_{\tau i}|^{2} = 5 \times
10^{-3} \gg |R^{}_{e i}|^{2}$ (for $i=1,2$) for illustration.}
\begin{tabular}{|c|c|c|c|c|c|c|c|c|}
\hline ~ & \multicolumn{2}{c|}{Signal} &
\multicolumn{2}{c|}{$t\overline{t}$} &
\multicolumn{2}{c|}{$W^{\pm}W^{\pm}W^{\mp}$} & \multicolumn{2}{c|}{$W^{\pm}W^{\pm}jj$} \\
\hline ~  & ~~$\sigma$(fb)~~ & ~~~eff.~~~ & ~~$\sigma$(fb)~~ &
~~~eff.~~~ & ~~$\sigma$(fb)~~ & ~~~eff.~~~ & ~~$\sigma$(fb)~~ & ~~~eff.~~~ \\
\hline Basic cuts & $377.7$ & $-$ &$32.0$ &$-$ &$0.23$ & $-$ & $2.04$ & $-$ \\
\hline + $\slashed{E}^{}_{\rm T}$ cut & $243.4$ & $64.4\%$
&$4.98$ & $15.6\%$ & $0.017$ & $7.39\%$ & $0.18$ & $8.82\%$ \\
\hline
\end{tabular}
\end{center}
\end{table}

Our numerical results obtained by inputting some typical values make
clear the main features of the collider signals for more than one
heavy Majorana neutrinos at the LHC. A systematic analysis of the
parameter space is desirable and can be done in a similar way. It is
worthwhile to note that a detailed study of the couplings
$R^{}_{\alpha i}$ has been performed in Ref. \cite{Pavel}, where
$R^{}_{\alpha i}$ are reconstructed from low-energy neutrino mixing
parameters and heavy Majorana neutrino masses. It has been found
that the collider signals are closely correlated with the mass
hierarchies of light neutrinos, and also the mass spectra of heavy
Majorana neutrinos \cite{Pavel}. From the above discussions, we can
conclude that two heavy Majorana neutrinos may induce significant
and constructive interference in the total cross section of the
same-sign dilepton signals. Moreover, the angular correlation
between the final charged leptons and the invariant mass
reconstruction from them can provide important information, which
may be used to distinguish the LNV signals induced by heavy Majorana
neutrinos from those by the doubly-charged Higgs bosons.

\section{Concluding Remarks}

Neutrino oscillation experiments have provided robust evidence that
neutrinos are massive. To explain tiny neutrino masses, one should
go beyond the SM. Therefore, at the high energy frontier to be
explored by the LHC, we also hope to gain some hints on or even to
pin down the mechanism of neutrino mass generation. To be specific,
we can test the popular seesaw models of neutrino masses at the LHC.

In the canonical seesaw model, the structural cancellation condition
$M^{}_{\rm D} M^{-1}_{\rm R} M^T_{\rm D} = {\bf 0}$ is required to
guarantee that the charged-current interactions of heavy Majorana
neutrinos are significant, while their masses can be as low as
several hundred GeV. Thus the heavy Majorana neutrinos can be
discovered at the LHC via the same-sign dilepton signals $pp \to
l^\pm_\alpha l^\pm_\beta jj$ (for $\alpha, \beta = e, \mu, \tau$).
Starting from the seesaw formula $M^{}_\nu = M^{}_{\rm L} -
M^{}_{\rm D} M^{-1}_{\rm R} M^T_{\rm D}$ and examining the interplay
between the two terms on the right-hand side, we have classified the
testable type-(I+II) seesaw model, where both heavy Majorana
neutrinos and a triplet scalar are introduced. An intriguing
type-(I+II) seesaw model with $M^{}_{\rm D} M^{-1}_{\rm R} M^T_{\rm
D} = {\bf 0}$ and then $M^{}_\nu = M^{}_{\rm L}$, which is achieved
by the discrete ${\rm A^{}_4 \times Z^{}_2}$ symmetry, has been
discussed in some detail. It has been found that (a) the non-unitary
neutrino mixing matrix is of the tri-bimaximal pattern in the
leading-order approximation; (b) the heavy Majorana neutrinos are
degenerate in mass, so the light neutrino masses are rather stable
against the radiative corrections. This scenario is a typical
example of the realistic type-(I+II) seesaw model with more than one
heavy Majorana neutrinos \cite{Choubey}.

Furthermore, we have calculated the cross section of the same-sign
dilepton signals $pp \to l^\pm_\alpha N^{}_i \to l^\pm_\alpha
l^\pm_\beta jj$ (for $i = 1, 2$ and $\alpha, \beta = e, \mu, \tau$)
in the minimal type-(I+II) seesaw model. The angular distribution as
well as invarinat mass distribution of final charged leptons can be
used to discriminate the signatures induced by heavy Majorana
neutrinos from those by the doubly-charged Higgs bosons. Making use
of some kinematic cuts, we have demonstrated that the SM backgrounds
can be rendered to be extremely small. It is worthwhile to stress
that the constructive interference of the contributions from two
heavy Majorana neutrinos may enhance the total signal cross section
by a factor up to four. In addition, we put forward an efficient
method to reconstruct the masses of heavy Majorana neutrinos. These
distinct features of our scenario may show up in the forthcoming
CERN LHC, so we hope that the LHC will shed some light on the
dynamics of neutrino mass generation in the near future.

\appendix

\section{The Calculation of Cross Section}

In this appendix, we show some details of the calculation of the
total cross section for the processes $pp \to l^\pm_\alpha
l^\pm_\beta j j$, which are induced by two heavy Majorana neutrinos
$N^{}_1$ and $N^{}_2$. First, we write down the Feynman amplitude
for the partonic process in Eq. (\ref{eq:process}), and the squared
matrix element is given in Eq. (\ref{eq:amplitude}). The relevant
functions quoted therein can be cast into a compact form by defining
a scalar function
\begin{eqnarray}
D^{}_X(p^2)=\frac{1}{p^2 - M^2_X + i M^{}_X \Gamma^{}_X} \;
\end{eqnarray}
for the unstable particle $X$ with mass $M^{}_X$ and total decay
width $\Gamma^{}_X$. For instance, the function $D^{}_W(p^2)$ for
the charged gauge bosons $W^\pm$ can be obtained by inputting
$M^{}_W = 80.398~{\rm GeV}$ and $\Gamma^{}_W = 2.141~{\rm GeV}$
\cite{PDG}. Likewise for the heavy Majorana neutrinos $N^{}_1$ and
$N^{}_2$. Therefore, the functions ${\cal F}^{}_i$ (for $i = 1, 2$)
and ${\cal G}$ are given by
\begin{eqnarray}
{\cal F}^{}_i  &=& {\rm
Re}\left[D^{}_i(k^2_3)D^*_i(k^2_4)\right]\left(p^{}_1 \cdot
p^{}_5\right) \left(p^{}_3\cdot p^{}_4\right) + {\rm
Im}\left[D^{}_i(k^2_3)D^*_i(k^2_4)\right]\varepsilon^{}_{\mu \nu
\lambda \delta} p^{\mu}_1 p^{\nu}_3 p^{\lambda}_4 p^{\delta}_5 \nonumber \\
&& + \left\{|D^{}_{i}(k^2_4)|^2 - {\rm
Re}\left[D^{}_i(k^2_3)D^*_i(k^2_4)\right] \right\} (p^{}_1\cdot
p^{}_4)(p^{}_3\cdot p^{}_5) \nonumber \\
&& + \left\{|D^{}_{i}(k^2_3)|^2 - {\rm
Re}\left[D^{}_i(k^2_3)D^*_i(k^2_4)\right] \right\} (p^{}_1\cdot
p^{}_3)(p^{}_4\cdot p^{}_5) \;
\end{eqnarray}
and
\begin{eqnarray}
{\cal G} &=& \left[2D^{}_1(k^2_3) D^*_2(k^2_3) -
D^{}_1(k^2_4)D_2^*(k^2_3) - D_1(k^2_3)D_2^*(k^2_4)\right] (p^{}_1
\cdot p^{}_3)(p^{}_4 \cdot p^{}_5) \nonumber \\
&& + \left[2D^{}_1(k^2_4) D^*_2(k^2_4) - D^{}_1(k^2_4) D^*_2(k^2_3)
- D^{}_1(k^2_3) D^*_2(k^2_4)\right] (p^{}_1 \cdot p^{}_4)(p^{}_3
\cdot p^{}_5) \nonumber \\
&& + \left[ D^{}_1(k^2_3) D_2^*(k^2_4) + D^{}_1(k^2_4) D^*_2(k^2_3)\right]
(p^{}_1 \cdot p^{}_5) (p^{}_3 \cdot p^{}_4) \nonumber \\
&&- \left[ D^{}_1(k^2_3) D_2^*(k^2_4) - D^{}_1(k^2_4)
D^*_2(k^2_3)\right] i\varepsilon^{}_{\mu \nu \lambda \delta}
p^{\mu}_1 p^{\nu}_3 p^{\lambda}_4 p^{\delta}_5 \;
\end{eqnarray}
with $k^{}_3 \equiv p^{}_1 + p^{}_2 - p^{}_3$ and $k^{}_4 \equiv
p^{}_1 + p^{}_2 - p^{}_4$. Starting with Eq. (\ref{eq:amplitude}),
we can calculate the cross section at the parton level, then the
total cross section by using Eq. (\ref{eq:crosssection}) and also
the parton distribution functions CTEQ6L1 \cite{CTEQ6}.

\section*{Acknowledgment}

The authors would like to thank Prof. Z.Z. Xing for his stimulating
discussions, which initiate the present work. This work was
supported in part by the National Natural Science Foundation of
China.

\end{document}